\documentclass[aps,prl,final,notitlepage,sort&compress,twocolumn]{elsarticle}
\newcommand{\degr}{\ensuremath{^\circ}}
\usepackage[pdfauthor={Tim Maxwell}, pdfstartview={FitB}, dvipdfmx,hidelinks]{hyperref}
\usepackage{amsmath}
\usepackage{amssymb}
\usepackage{graphicx}
\usepackage{subfigure}
\usepackage{appendix}
\usepackage{txfonts}
\begin{document}

\title{Synchronization and Characterization of an Ultra-Short Laser for Photoemission and Electron-Beam Diagnostics Studies at a Radio Frequency Photoinjector}

\author[niu]{Timothy Maxwell}
\author[ad]{Jinhao Ruan}
\author[niu,apc]{Philippe Piot}
\author[ad]{Alex Lumpkin}
\address[niu]{Northern Illinois Center for Accelerator \& Detector Development and Department of Physics, \\
Northern Illinois University, DeKalb, IL  60115, USA}
\address[ad]{Accelerator Division, Fermi National Accelerator Laboratory, Batavia, IL  60510, USA}
\address[apc]{Accelerator Physics Center, Fermi National Accelerator Laboratory, Batavia, IL  60510, USA}
\date{December 16, 2011}

\begin{abstract}
A commercially-available titanium-sapphire laser system has recently been installed at the Fermilab A0 photoinjector laboratory in support of photoemission and electron beam diagnostics studies.  The laser system is synchronized to both the 1.3-GHz master oscillator and a 1-Hz signal use to trigger the radiofrequency system and instrumentation acquisition.  The synchronization scheme and performance are detailed.  Long-term temporal and intensity drifts are identified and actively suppressed to within 1 ps and 1.5\%, respectively. Measurement and optimization of the laser's temporal profile are accomplished using frequency-resolved optical gating.
\end{abstract}
\maketitle

\section{Introduction}
The A0 photoinjector (A0PI) facility \cite{Carneiro2005} at Fermilab has provided electron beam in support of a variety of advanced accelerator R\&D experiments over the last decade. The facility includes a picosecond neodymium-doped yttrium lithium fluoride (Nd:YLF) laser system \cite{Li} used to photoemit the electron bunches from a cesium-telluride (Cs$_2$Te) photocathode. Recently, an ultra-short, titanium-sapphire (Ti:sapph) laser system was installed to enable the formation of ellipsoidal bunches in the blow-out regime \cite{ellipsebunch, uclaellipse, Musumeci}, and the development of novel diagnostics utilizing electro-optic spectral decoding (EOSD) ~\cite{specencode, Jiang1998}.

For ellipsoidal bunch generation experiments, the amplified 800-nm infrared (IR) laser output is tripled to the ultraviolet (UV, 266 nm) as required for photoemission from a Cs$_2$Te photocathode.  The UV output must be well-synchronized to the L-band RF gun and of sub-picosecond duration.

Experiments in EOSD encode the terahertz-domain transient electromagnetic field (either radiation or velocity field) from the electron beam onto a stretched, broadband laser probe pulse for longitudinal bunch distribution monitoring.  To consistently see the beam's full longitudinal profile, one must minimally chirp the pulse to the electron bunch length plus some additional tolerance for shot-to-shot temporal fluctuations.  Reduction of these fluctuations to a maximum of 1 ps is required to maintain sufficient time resolution~\cite{optimumlength}.

In this paper we discuss the installation of the Ti:sapph laser system and verify that it meets the above requirements.  We begin with a brief description of the layouts of the accelerator and Ti:sapph laser system.  In following sections we describe the scheme used to synchronize the new seed laser and amplifier.  This includes discussion on accounting for potential intensity instabilities introduced by unstable triggering of the 1-kHz pulse selection in the Ti:sapph regenerative amplifier (regen). As EOSD is to be performed on beam generated using the existing Nd:YLF laser system, good cross-synchronization of the two lasers is as important as timing to the accelerator.  Temporal stability of both laser systems are therefore examined. We conclude with measurement and optimization of the laser pulse length by frequency-resolved optical gating (FROG).

\section{Experimental setup}
In this section we provide a summary of the photoinjector layout including details on the synchronized streak camera that was used for these experiments.  The optical layout of the Ti:sapph laser is then presented with details on its specifications.

\subsection{Accelerator Layout}\label{sec:acceleratorlayout}
\begin{figure}[!hb]
\centering
\includegraphics[width=3.1156in]{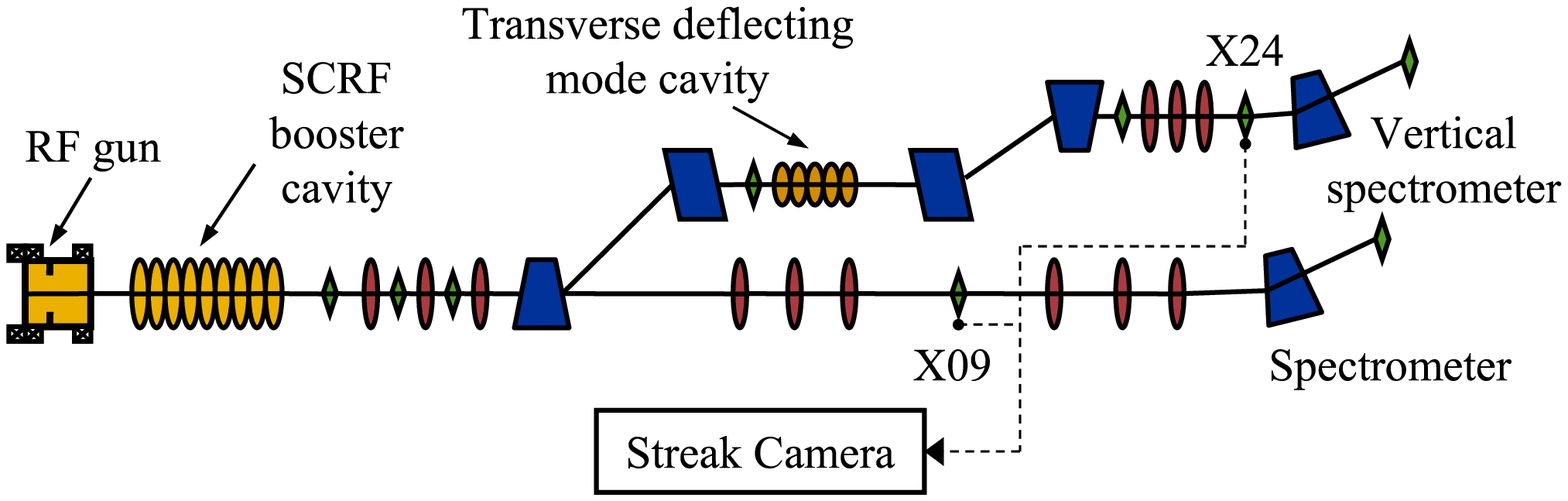}
\caption{\label{fig:a0beamline}Top view of the A0 photoinjector setup. The red ellipses and blue trapezoids are quadrupole and dipole magnets, respectively. The OTR stations used for temporal jitter investigation are labeled as X09 and X24 with optical paths to the streak camera shown.}
\end{figure}

Shown schematically in Fig.~\ref{fig:a0beamline}, the photoinjector utilizes a Cs$_{2}$Te photocathode in a 1-1/2 cell L-band, 1.3-GHz RF gun as electron source.  The photocathode is typically driven by the amplified, frequency quadrupled output of the existing Nd:YLF laser system~\cite{Li}.  Downstream of the gun,  a 1.3-GHz, 9-cell superconducting RF (SCRF) booster cavity accelerates the $\sim 1$~nC beam up to 16 MeV. The beamline also includes quadrupole and dipoles magnets necessary to control the beam's trajectory and size. The beam then propagates down either the ``straight-ahead"' beamline for diagnostics and user studies or the ``emittance exchange beamline"~\cite{Ruan,Sun}.  As shown in Fig.~\ref{fig:a0beamline}, the latter consists of a 5-cell transverse deflecting-mode cavity\cite{tim5cell} flanked between two doglegs.  The RF gun, 9-cell SCRF cavity and 3.9-GHz deflecting mode cavity are all synchronized to the 1.3-GHz RF master oscillator.

The accelerator incorporates transverse and longitudinal phase-space diagnostic stations. The diagnostics pertinent to the experiments presented in this paper are shown as X09 and X24 in Fig.~\ref{fig:a0beamline}. Both stations generate backward optical transition radiation (OTR) using the conventional arrangement. The OTR from either cross can be imaged to the entrance slit of a streak camera for bunch length measurement.

The streak camera is equipped with synchroscan and phase lock loop (PLL) electronics to maintain synchronization with the 81.25-MHz subharmonic of the A0PI master oscillator~\cite{streak}.  This can be operated at four sweep rates to adjust temporal range and resolution.  In its fastest sweep range, referred to as range 1, the single-sweep resolution is 540 fs RMS with the next-fastest sweep rate (range 2) having a temporal resolution of 2.5 ps.

As the streak camera sweep unit is phase locked to the RF, it is also used in this analysis to perform synchronization measurements by tracking the centroid of streak images.  The standard deviation of shot-to-shot changes in streak image position from tracking a similarly phase-locked laser pulse train is typically found to be $\sim$1 pixel.  This corresponds to 320~fs in sweep range 1 and 1.5 ps in range 2.

\subsection{Ti:Sapph Laser Layout}
The new Ti:sapph laser system is a commercially available \emph{Spitfire Pro XP} regen seeded by a \emph{Tsunami} oscillator.  These are respectively pumped by a 30 W, Q-switched \emph{Empower} laser and 5 W continuous-wave \emph{Millennia Pro} diode laser, respectively, with both pumps operating at 532~nm.  The full system, produced by Newport Corporation, Spectra Physics division, produce 800-nm pulses with output parameters summarized in Table~\ref{tab:laserparams}.  The system is supplemented with a \emph{Dazzler} produced by Fastlite~\cite{aopdf}, an acousto-optic programmable dispersive filter (AOPDF), to allow for temporal pulse shaping of the IR pulse.
\begin{table}[h]
\centering
\caption{Optimized Ti:sapph laser parameters}
\label{tab:laserparams}
\begin{tabular}{p{1.6in}|p{0.9in}}
  \cline{1-2}
  Oscillator center wavelength & 800 nm \\
  Oscillator repetition rate & 81.25 MHz \\
  Oscillator pulse energy & 14.5 nJ \\
  Oscillator max bandwidth & 15 nm (FWHM) \\
  Amplified repetition rate & 1 kHz \\
  Amplified pulse energy & 3 mJ \\
  Amplified max bandwidth & 12 nm (FWHM) \\
  \hline
\end{tabular}
\end{table}

\begin{figure}[htpb]
\centering
\includegraphics[width=3.24542in]{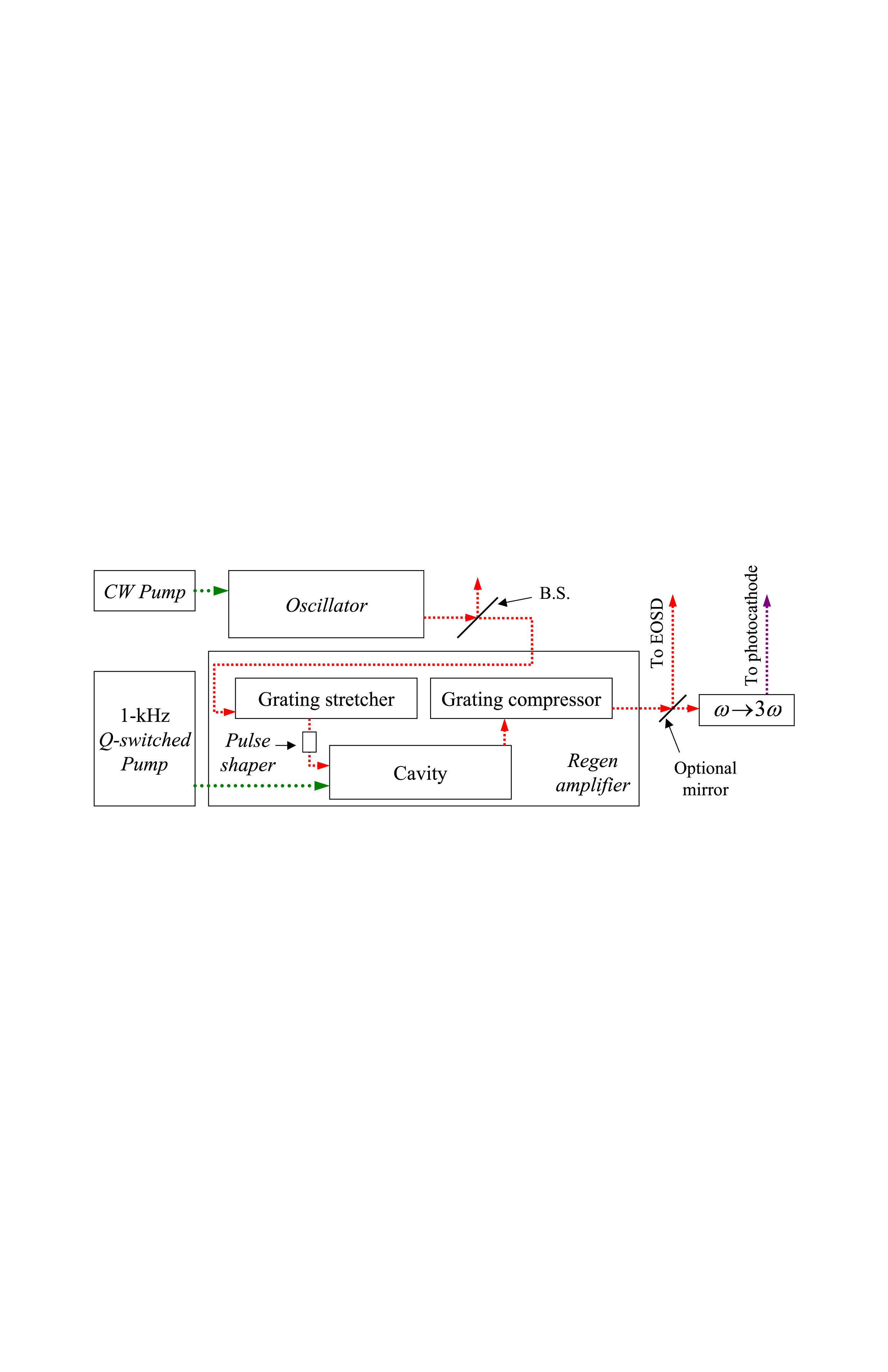}
\caption{\label{fig:opticallayout}Schematic of the modified optical layout for the Ti:sapph laser system including oscillator pick-off beam splitter and pulse shaper.}
\end{figure}

The layout of the laser system is shown in Fig.~\ref{fig:opticallayout}.  The seed laser output passes a 75\% reflective, 25\% transmissive pellicle beam splitter (BS).  The reflected beam is used for optional timing feedback (see Sec.~\ref{sec:secondarypll}) and other diagnostics.  The transmitted beam is passed as the amplifier seed.


The amplified IR output was chosen for EOSD where a strongly chirped IR laser pulse is required. The amplifier's grating compressor in combination with the \emph{Dazzler} can easily produce the required ps-scale chirp.  Further, EOSD occurs on a single-shot basis with the signal from the beam modulating the spectrum of the laser pulse~\cite{specencode, Jiang1998}.  The regen selects only a 1-kHz pulse train (or less with gating) for amplification.  This provides suppression of the seed laser's 81.25-MHz repetition rate to the accelerator's 1 Hz, mitigating concerns about the EOSD optical spectrometer camera integrating pulses not modulated by EOSD.

\begin{figure}[htpb]
\centering
\includegraphics[width=3.24542in]{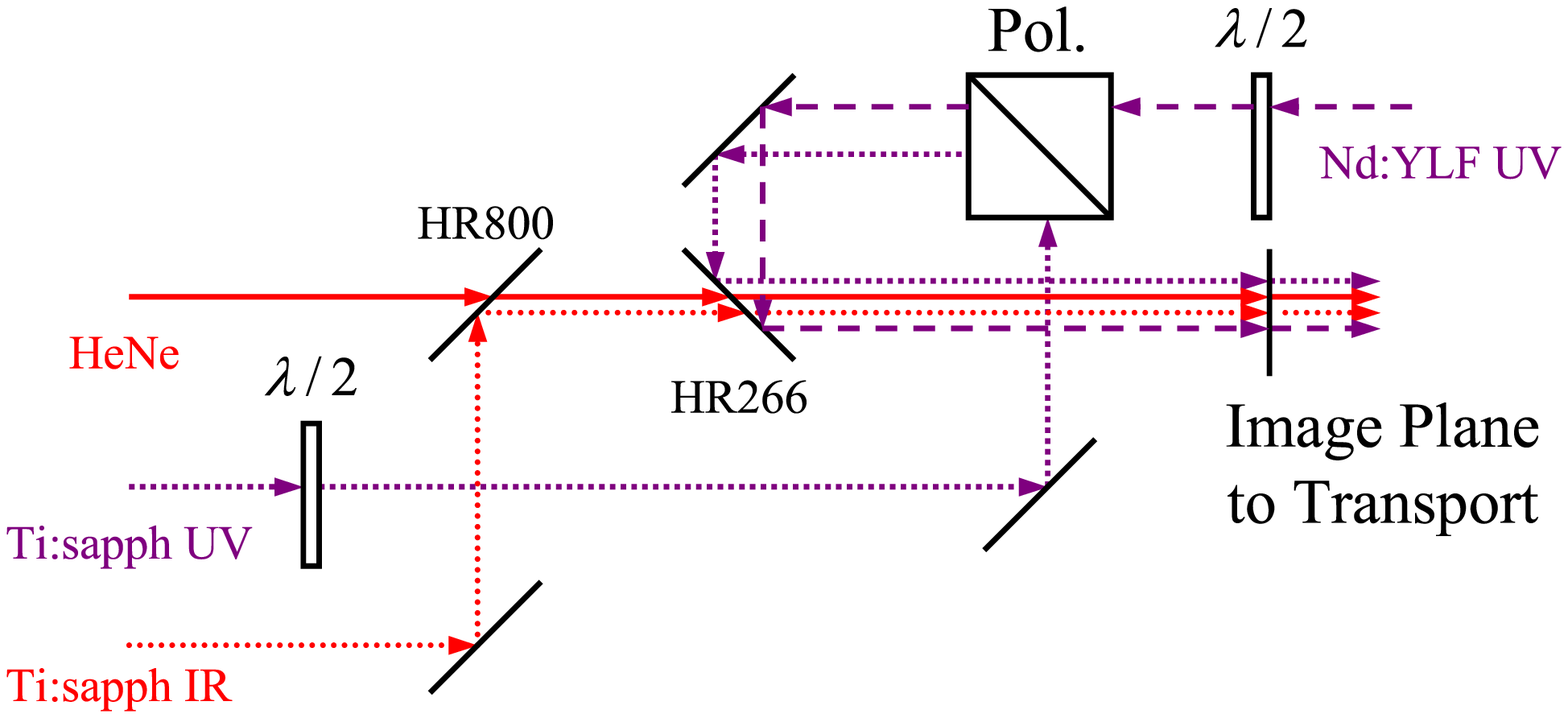}
\caption{\label{fig:combiner}Optical layout used to combine the Ti:sapph UV, Ti:sapph IR, Nd:YLF UV and HeNe alignment laser before transport to the accelerator tunnel.  Polarization of the two UV beams are controlled independently using the two half-wave plates and combined in a UV polarizing cube.  A high-reflectivity (HR) 800-nm mirror combines the IR with the HeNe (632 nm) alignment lasers then joined with the UV beams upon transmission through the HR266 mirror.}
\end{figure}

To accommodate concurrent delivery of the Ti:sapph and existing Nd:YLF and HeNe alignment lasers to the accelerator vault, they are combined before transport; see Fig.~\ref{fig:combiner}.  The combined beams are then sent into an optical transport line to deliver pulses from the A0PI laser lab to an optical breadboard at the photocathode in the accelerator tunnel 20 m away.  The transport line consists of five mirrors with a double high-reflectivity (HR) coating at 800 nm and 266 nm and with a 5-m imaging lens at its midpoint.

\section{Synchronization with the photoinjector}
Synchronizing the Ti:sapph output to the accelerator requires management of two time scales.  The first is ns-scale synchronization of the regen amplifier to the photoinjector 1-Hz event that triggers the RF pulse forming network and instrumentation.  The regen selects a 1-kHz train of pulses from the 81.25-MHz seed train with the ability to gate any subset of this train.  A 1-kHz trigger must be provided to select these pulses in sync with the 1-Hz event so that both references select the same RF cycle from their higher harmonics.  This ensures regular timing of the laser with respect to the gun RF pulse and data acquisition from shot to shot.

The second time scale is fs-scale synchronization of the seed laser 81.25-MHz pulse train to the 1.3-GHz master oscillator that drives the photoinjector RF.  This drives the stability of the launch phase at the photocathode when operating in UV drive laser mode and beam-probe laser jitter in the context of EOSD.

We discuss the fine temporal stability in the usual parameters of jitter and drift.  Jitter refers to the spread of shot-to-shot fluctuations in the output time difference of the seed laser with respect to some reference signal.  This is typically less than 1 ps.  Drift is the change in the mean time difference from a reference signal over extended periods of the order ps/hour.


\subsection{Coarse triggering of the Ti:sapph amplifier}
With the Ti:sapph regen's fine synchronization driven primarily by that of the seed laser, we first look at coarse timing of its 1-kHz repetition rate to the 1-Hz A0PI clock.  The 1-kHz trigger is used by the \emph{Dazzler} and \emph{Spitfire} timing electronics to determine which pulses in the 81.25-MHz seed laser pulse train to capture for shaping and amplification.

As the gun RF is fed by an RF macropulse with 400 $\muup$s maximum duration and some instrumentation (e.g. beam position monitors) only diagnose the first bunch in each macropulse, only one pulse of a given 1000-pulse regen cycle is relevant to beam physics experiments at the photoinjector.

\begin{figure}[htb]
\centering
\includegraphics[width=3.18051in]{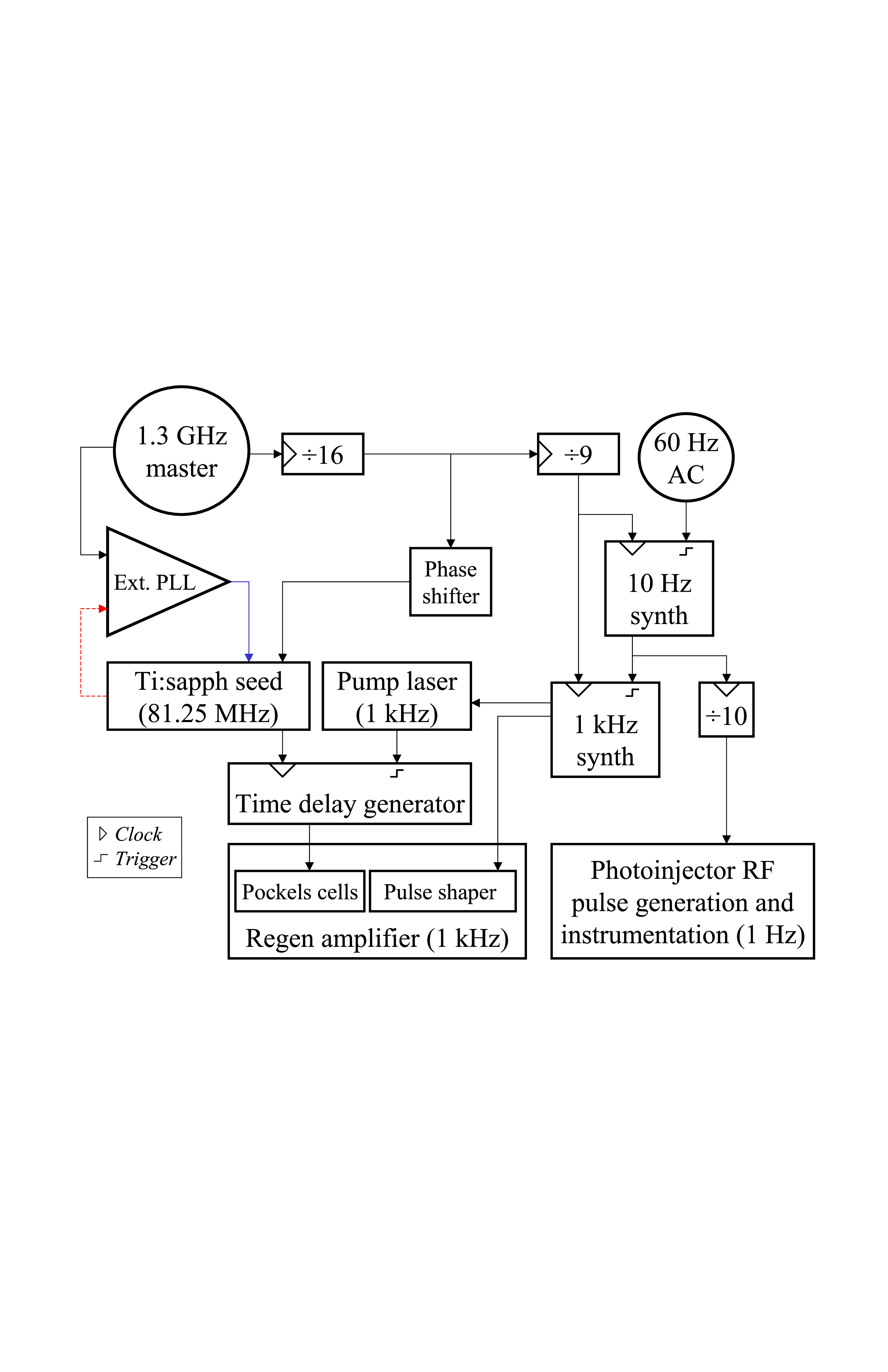}
\caption{\label{fig:timinglayout}Block diagram of timing scheme used to synchronize the Ti:sapph laser system to the photoinjector.}
\end{figure}

A block diagram of the full synchronization scheme is shown in Fig~\ref{fig:timinglayout}.  The 81.25-MHz seed laser output is internally synchronized to the $16^{th}$ subharmonic of the 1.3-GHz master oscillator.  Where additional fine temporal phase monitoring and control are needed, a secondary, external PLL has been added.  Details on this are described later in Sec.~\ref{sec:secondarypll} with the secondary, external PLL illustrated in Fig.~\ref{fig:feedbackloop}.

For slow trigger generation the 81.25-MHz sub-master is further divided down to 9.028~MHz and used as the clock for two digital counter synthesizers integrated on a system of field programmable gate arrays (FPGA).  The first of these is the existing 10-Hz signal generator and count-down dividers providing 5-, 2- and 1-Hz signals.  The signal generator selects the 10-Hz to be in phase with the building's 60-Hz AC.
%

The 1-kHz trigger is generated by upsampling of the slower 10-Hz clock.  This 1-kHz signal fires the regen's Q-switched pump laser, the regen pulse-picking Pockels cells, and the \emph{Dazzler} pulse shaper.

A $\muup$s-scale instability is introduced by this synchronization of the 10 Hz to the 60-Hz main line.  As will be demonstrated throughout this section, this adds a number of complications to the triggering system with regards to 1-kHz trigger generation, \emph{Dazzler} timing requirements, and the stability of the regen pump laser.

The \emph{Dazzler} diffracts the shaped laser pulse train into the regen cavity for amplification by loading a 33 $\muup$s-duration acoustic waveform into the AOPDF.  The electronics to load the waveform must be triggered 25 $\muup$s before the regen system.  Therefore the 1-kHz trigger must provide two output channels with independent time delays to allow the \emph{Dazzler} to be fired in advance of the regen.  Where a 1-kHz clock stable to the microsecond is available, this becomes unnecessary as the \emph{Dazzler} can instead be configured to trigger on delay of the previous pulse in the train.

For the Pockels cells, the 1-kHz event is synchronized by the time delay generator (TDG) to be locked with the 81.25-MHz signal of the seed laser for pulse selection. A 1-Hz, 1 ms-duration logical signal is then used to gate only the 1-Hz pulses on-time with the gun RF pulse.

The coarse, 1-kHz trigger signal has been derived passively using a signal generator as well as actively using a preferred custom trigger synthesizer built on an Altera (part no. EPF10K40RC208-3) FPGA for our purposes.

\subsubsection{1-kHz burst generation}
In the simplest case, an externally triggered signal generator operating in burst mode can produce a train of 1-kHz pulses fired at each 1-Hz event to produce a steady signal.  The burst generator is set to fire a fixed number of pulses at a fixed frequency of approximately 1-kHz sufficient to fill the time between slow trigger events.

If at any time a trigger event occurs before a previous pulse train completes firing, that event is ignored by the signal generator causing it to remain low until the following macropulse event.  This gap in triggering will cause an under-frequency protection fault in the amplifier pump laser, disabling the regen.  Therefore for uninterrupted operation, one must choose the pulse generator's macropulse duration (product of number of cycles and repetition rate) to be smaller than the shortest possible trigger period.

As stated above, this becomes difficult in our case due to the slow frequency fluctuations driven by the dependence of the low repetition rate clocks on the 60-Hz main voltage.  Twenty-four hour recording of the period for these, as accumulated with an oscilloscope, reveal that the 1-Hz clock period can vary as much as $\pm$2.5 ms from the nominal 1 s.  If the macropulse from the signal generator is set to a 997.5 ms duration there can then be as much as 5 ms of dead time in the limiting case.  This is again sufficient to generate an under-frequency fault in the regen pump laser.

We instead upsample the 10-Hz signal which accumulates a smaller phase difference from 60-Hz frequency deviations between clock resets.  Measurement over 8-hours shows the 10-Hz period varying up to only 120 $\muup$s from the nominal 100 ms with a root-mean-squared (RMS) deviation of 30 $\muup$s.  This allows for a burst generator operating with 100 pulses at 1.0012 kHz to have a maximum dead time of 1.24 ms for any given shot, sufficient to maintain laser operation.

\subsubsection{Intensity dependence on unstable triggering}
Using the above burst generator as the 1-kHz trigger means allowing the period before the first cycle in each burst to vary several tens of microseconds.  This unstable repetition rate can adversely affect the \emph{Empower} and, therefore, amplified IR output.

The explanation for the transient response from a sudden change in repetition rate can be understood in terms of the population inversion seen from one shot to the next~\cite{lasers}.  After a pulse is emitted there is some residual inversion left in the Q-switched cavity.  This will then build back up over the pumping period to a new initial inversion before the Q-switch is reopened for pulsing, dropping the inversion down to a new residual.

For a constant pumping period, the initial and residual inversions will reach a steady state.  However, a sudden change in this period will temporarily disrupt this as some additional (or lesser) inversion is built up before the next pulse is emitted.

The effect of \emph{Empower} repetition rate on the output power and pulse shape of the Q-switched laser can be modeled with detailed knowledge of cavity properties such as upper state lifetime of the lasing medium, cavity decay rate, inversion threshold and pumping rate~\cite{lasers, qshape}.  The simple numerical model for a repetitively Q-switched laser suggested in~\cite{lasers} illustrates that for parameters typical of a Nd:YLF system, reasonably steady-state output is reached within the first 2--3 cycles after a change in repetition rate.

The Q-switched build-up time $\tau_n$ associated with any $n$-th pulse will be related to the preceding build-up time $\tau_{n-1}$.  For our burst generator configuration we have
\begin{equation}\label{eq:taupiece}
\tau_n = \begin{cases}
  \tau_{kHz} & \text{for $n \ne N$} \\
  \tau_{Trig} - (N - 1) \tau_{kHz} & \text{for $n = N$}\end{cases}
\end{equation}
Where $\tau_{kHz} \approx 1 $ms is the burst generator period, $N = 100$ is the number of pulses in a macropulse, and $\tau_{Trig} \ge N \tau_{kHz} = 100$ ms is the changing, external 10-Hz trigger period.  With the repetition rate (build-up time) disrupted only on every $N$-th cycle, the modeling suggests that only the following $n = 1$ cycle has its pulse energy disturbed greater than a fraction of a percent.

Characterization of how the single-cycle triggering impacts laser performance was done empirically.  To control the disruption, two signal generators are used.  The first is used as the variable 10-Hz or slower ($\tau_{Trig} \ge $ 100 ms) event while the second acts as the burst generator firing $N$ = 100 pulses with a fixed $\tau_{kHz} = 999.1$ $\muup$s spacing externally triggered by the slow generator.

\begin{figure}[htb]
\centering
\includegraphics[width=3.05069in]{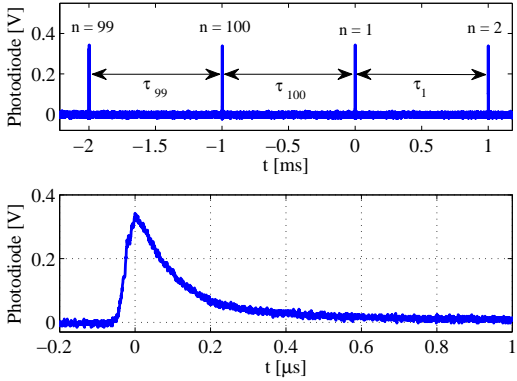}
\caption{\label{fig:4mssample}Example sweep measuring regen pump laser intensity dependence on the timing disruption introduced by 1-kHz, 100 pulse burst generator being triggered by a $>$10 Hz event (top).  On each sweep, the amplitude and area of the zoomed traces (bottom) of each of the $n = 99, 100, 1$ and $2$ pulses in the macropulse are recorded.}
\end{figure}

A fast, 1-ns photodiode is connected to a 2-GHz oscilloscope monitoring the \emph{Empower} output and triggered on the 10-Hz event.  In this way we observe the output of the first pulse after the discontinuity in timing as well as its nearest neighbors with a 4-ms sample taken at each shot monitoring the $n =  99, 100, 1$ and $2$ pulses of each macropulse train.  With a zoomed trace on each of the four pulses, their amplitude and area can be averaged over several shots to record the peak and total pulse energy of each for a given $\tau_{Trig}$ and the associated disrupted $\tau_N$, as illustrated in Fig.~\ref{fig:4mssample}.

\begin{figure}[tb]
\centering
\includegraphics[width=2.72615in]{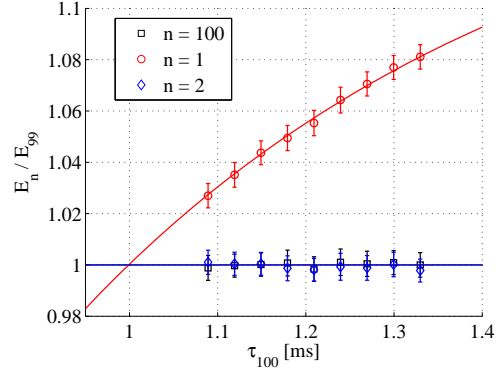}
\caption{\label{fig:empowerplot} Relative deviation of $n$-th \emph{Empower} pulse energies within the 100-pulse train as a function of $\tau_{100}$ from using the burst generator setup.  For all other $n \ne 100$ in the train, $\tau_n = $ 999.1 $\muup$s (see Fig.~\ref{fig:4mssample}).}
\end{figure}

We expect the output to be steady as it approaches $n = 100$ and to quickly recover after the disrupted $n = 1$ pulse.  The $n = 100, 1$ and $2$ pulses energies $E_n$ are normalized to the energy $E_{99}$ of the 99-th pulse in each train to monitor relative changes to the steady state output.  This data is shown in Fig.~\ref{fig:empowerplot} as a function of the disrupted period $\tau_{100}$ from varying $\tau_{Trig}$ (Eq.~\ref{eq:taupiece}).  To extrapolate to values outside the range measured, a trend line based on cavity parameters roughly estimated using the measured pulse shape on a fast photodiode~\cite{qshape} combined with a fit to this data using the numerical models in~\cite{lasers} is shown.

We note that the statistical error bars shown in Fig.~\ref{fig:empowerplot} are found to be driven primarily by digital sampling error from the use of the lower time resolution zoomed traces and not indicative of actual energy fluctuations.  As evidence of this, contracting the sampling region to cover only one pulse reduced this error to less than 1\%.

Fig.~\ref{fig:empowerplot} clearly demonstrates the expected behavior with the energy of the first pulse $E_{1}$ deviating from the steady-state output at a rate of 0.023\% per $\muup$s change in $\tau_{100}$.  As the 10-Hz photoinjector trigger (Fig.~\ref{fig:timinglayout}) demands operating with as much as 240 $\muup$s variation in $\tau_{Trig}$, $E_1$ can be expected to vary as much as 3--8\% from nominal.

As expected, however, the output quickly recovers by the second $n = 2$ pulse.  Within error bars there are no observable deviations in output with the fitted model suggesting relative fluctuations are limited to $< 2\times10^{-4}\%$ over the range shown.

\begin{figure}[tbp]
\centering
\includegraphics[width=2.72615in]{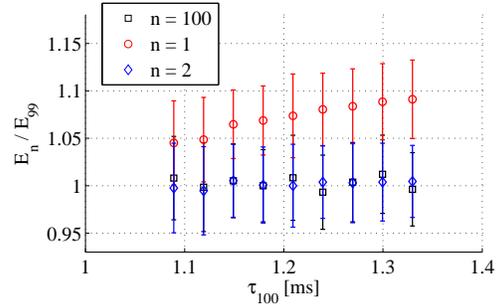}
\caption{\label{fig:spitfireplot}Measured relative deviation of $n$-th \emph{Spitfire} pulse energies within the 100-pulse train as a function of $\tau_{100}$ from using the burst generator setup.}
\end{figure}

To verify that this \emph{Empower} behavior translates to an observable effect on the regen amplifier's output, the same measurement was carried out for the \emph{Spitfire} with data shown in Fig.~\ref{fig:spitfireplot}.  As the integrated output of the 1-ns photodiode is still being used to estimate pulse energy for the 100-fs amplified pulses, sampling error becomes much larger and the linearity of the diode response is questionable and not estimated here.  In spite of this, the larger pump energies for the $n = 1$ pulse are resolved as larger amplified pulses, though changes in $E_2$ are not observed within measurement limits.

From these we see that the first pulse in every burst will see unwanted fluctuations while the second appears acceptably stable.  Therefore for both the EOSD probe pulse and UV drive laser pulse generation, a delayed 1-ms duration, 1-Hz gating signal is sent to the TDG to select only the second pulse.  With the 1-kHz signal generator delayed 0.5 ms from the 10-Hz signal and the gate delayed 1.5 ms, the second, more stable laser pulse is selected to be synchronous with the 1-Hz photoinjector clock for use with accelerator experiments.  Choosing a still-later pulse in the train may be generally preferred, however the 1-Hz master event for the photoinjector is delayed less than 2~ms from the 10-Hz signal, not allowing sufficient time to choose anything later than the second.

\subsubsection{1-kHz trigger synthesizer}
Using the above trigger generator configuration has been demonstrated to provide reasonable triggering for the regen laser system.  As a final solution, the FPGA-based active trigger synthesizer was also developed to further improve reliability.  Based on several built-in counters, it includes three primary improvements.

First, unlike an ordinary burst generator, the FPGA is ready to fire the next burst up to one half of a micropulse period before (or after) the end of the previous 100-cycle macropulse.  In this way the firing frequencies need not be chosen with an arbitrary offset of the last micropulse duration so that the mean value of $\tau_N$ is equal to $\tau_{n \ne N}$.

Second, if the 10-Hz trigger does not fire within this ${\pm}1/2$-cycle window, the synthesizer resumes continuous firing of its 1 ms-spaced pulse train while it waits for the next reset.  While this may result in a laser pulse not firing in sync with the 1-Hz of the accelerator on a given shot, it ensures that the \emph{Empower} firing rate stays within its specified 500 Hz -- 5 kHz range to prevent any under- or over-frequency protection faults from forcing shut down of the laser, regardless of the state of the 10-Hz signal which can suffer occasional disruptions.

Finally, it was observed that $\tau_{Trig}$ has a 30 $\muup$s RMS jitter with a slow drift of several tens of microseconds.  Adaptive feedback was therefore also included to reduce the maximum observed $\tau_{100}$ to the level of jitter in $\tau_{Trig}$ and ensure that subsequent shots regularly arrive within the 1-ms allowed window.  To this end, the synthesizer counts out the difference between the nominal $\tau_{Trig,0}$ = 100 ms and the actual time elapsed between successive slow trigger events.  For the next burst, this difference is absorbed into a small, fixed change $\epsilon_\tau$ in the micropulse period $\tau_{{\mu}P}$ = 1 ms over a variable number of cycles in the burst to make the macropulse approximately equal in duration to the previous 100-Hz period.

Explicitly, the adjusted micropulse period $\tau^{\prime}_{{\mu}P}$ is
\begin{equation}\label{eq:tauprime}
\tau^{\prime}_{{\mu}P} = \begin{cases}
\tau_{{\mu}P} + \epsilon_\tau & \text{for  $\tau_{Trig,prev} \ge \tau_{Trig,0}$} \\
\tau_{{\mu}P} - \epsilon_\tau & \text{for  $\tau_{Trig,prev} < \tau_{Trig,0}$} \end{cases}
\end{equation}
A dependence of the pump laser output on changes to the firing rate was demonstrated in the previous section. We have chosen $\epsilon_\tau$ to be 5 $\muup$s, just 0.5\% of the regular firing rate, to keep the effect introduced by the adaptive period negligible.  The number of cycles $N_{cor}$ the adjusted cycle is used is to correct the macropulse length is then,
\begin{equation}\label{eq:ncor}
N_{corr} = \text{int}\left(\frac{\left|\tau_{Trig,prev} - \tau_{Trig,0}\right|}{\epsilon_\tau}\right)
\end{equation}
where int($x$) is the nearest-integer rounding function.  The synthesizer will switch its repetition rate from $\tau_{{\mu}P}$ to $\tau^{\prime}_{{\mu}P}$ for $N_{corr}$ cycles so that the effective macropulse $\tau_{MP}$ length of the next burst will be
\begin{equation}\label{eq:mp}
\tau_{MP} = \left(N - N_{corr}\right) \tau_{{\mu}P} + N_{corr} \tau^{\prime}_{{\mu}P}
\end{equation}

As our application makes use of only the first or second cycle of any macropulse, the adapted period $\tau^{\prime}_{{\mu}P}$ isn't applied until later in the burst, running from cycles $n = 20$ to $n = 20 + N_{corr}$ to avoid any complication with the first few shots of interest.

\begin{figure}[htbp]
\centering
\includegraphics[width=3.05069in]{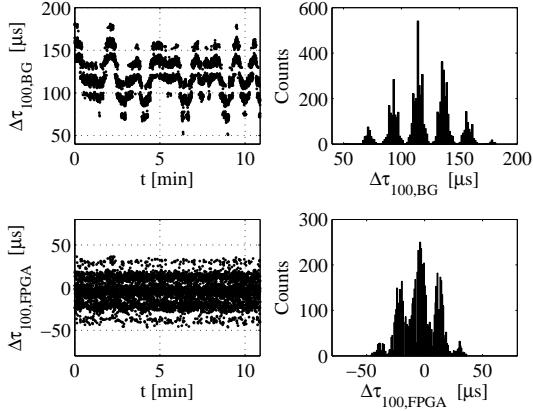}
\caption{\label{fig:fpga}Comparison of signals generated using a 1-kHz burst generator (top) and the FPGA synthesizer (bottom) showing the difference of the period of the final shot in the burst from the nominal micropulse period $\Delta\tau_{100}$ over time (left) and associated histograms (right).}
\end{figure}

We expect that the difference between the arrival of the next reset trigger from the 10-Hz signal will therefore regularly occur within $\epsilon_\tau \pm \langle \Delta\tau_{Trig}^2 \rangle^{1/2} \approx 35 \muup$s.  A measurement comparing output from the custom FPGA synthesizer to a burst generator appears in Fig.~\ref{fig:fpga} with both being triggered by the A0PI 10-Hz clock.
The FPGA reliably matches the end of one burst with the start of the next and without needing the large offset of the burst generator.  From Fig.~\ref{fig:fpga} we also see that the drift correlated to that of the slow trigger is effectively removed, reducing the maximum observed fluctuations.

Statistics accumulated over a 24-hour period show the FPGA output having a spread in the final period of just 15 $\muup$s RMS, an improvement over the 30 $\muup$s from the burst generator.  The maximum deviation of the period preceding the first shot in each macropulse $\tau_{100}$ is $\pm$40 $\muup$s for the FPGA over the same 24-hour period, in agreement with expectation and a good improvement over the $\pm$120 $\muup$s from the burst generator.

Using the FPGA synthesizer, the IR pulse energy standard deviation for the first pulse of each burst is found to be 3.2\% with no measurable intensity drift correlated to 10-Hz triggering drift.  This corresponds to a spread of 10.1\% for the UV output.  Using the more stable $n = 2$ pulse, this is reduced to 1.5\% in the IR and 4.0\% in the UV.

We find that with the adaptive correction, burst triggering dependably occurs on-time with each cycle.  With the FPGA synthesizer and second-cycle pulse selection, reliable operation of the amplifier is realized with no observed amplifier timing faults, missed shots, or reducible intensity fluctuations.

\subsection{Fine synchronization}
We now turn to the fine synchronization of both the added Ti:sapph system and existing Nd:YLF drive laser to the photoinjector RF.  Fine phase locking for both starts with the 16$^{th}$ sub-harmonic being counted down from the 1.3-GHz master oscillator and fed to the built-in locking electronics of the respective seed laser (Fig.~\ref{fig:timinglayout}).  These internal phase mixers compare the phase of the incoming 81.25-MHz signal with that of a photodiode mounted in the oscillator cavity.  The resulting difference voltage is used to drive a piezo-mounted intracavity mirror to maintain the phase lock.

The manufacturer specification for the \emph{Tsunami} oscillator's temporal jitter is 500 fs RMS.  To verify this a 20-GHz fast photodiode (Hamamatsu G4176-03 photodiode with Picosecond Pulse Labs 5545-107 bias tee) was used.  The signal after a custom 1.3-GHz bandpass, Q = 1000 cavity filter (Microwave Filter Co., Inc., 5MNSP-1.3/1.3) was measured using a signal source analyzer (Agilent E502B) consistently yielding an RMS jitter of better than 300 fs.  Similar measurements of the Nd:YLF system also yield a typical jitter of 300 fs RMS.

These are in good agreement with specification and with the less than 200-fs RMS jitter measured with the signal source analyzer from the 1.3-GHz master oscillator directly.

\subsubsection{Initial phase measurements}
To verify long-term stability we begin with measurements using the synchronized streak camera.  The laser at the photocathode surface is imaged onto the entrance slit of the streak camera.  The time of arrival is inferred from a fit of the digitized streak image.

Details on the resolution of the streak camera were given at the end of Sec.~\ref{sec:acceleratorlayout}.  However, as the PLL that maintains the streak camera's lock to the 81.25-MHz reference may also suffer from its own drift, a second technique was initially employed to verify the measured time of arrival using the launch phase sensitivity of the total charge emitted from the gun~\cite{chargemethod, chargemethod2}.

This ``charge technique'' is based on operating the RF gun at a low phase with respect to the photocathode drive laser.  In this regime, the emitted charge strongly depends on the phase between the laser and gun thereby providing a means to measure the jitter between the two systems.

Consider that some minimum gradient is needed to accelerate photoelectrons excited by the laser pulse to overcome the potential from their image charge at the photocathode.  Where the gun RF phase is set such that the sinusoidal gradient provided is lower than this critical value over the duration of the laser pulse, no charge is emitted.  As the phase of the RF is advanced, the gradient will increase until it exceeds this value and begins to accelerate the photoelectrons emitted by the head of the laser pulse.  Continuing to increase the phase will capture still more of the emitted electrons until the gradient is sufficiently high across the entire pulse and the full charge available by photoemission is accelerated. The total charge emitted can then be related to a partial integration of the laser's temporal profile in this regime~\cite{chargemethod}.

\begin{figure}[htpb]
\centering
\includegraphics[width=2.27179in]{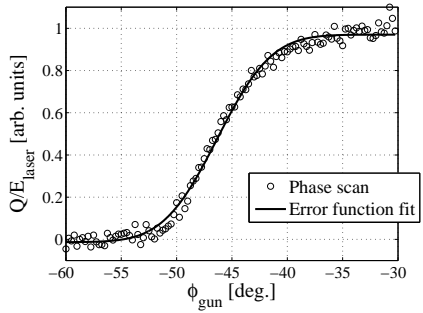}
\caption{\label{fig:phasescan}Scan of emitted charge normalized by laser pulse energy ($Q/E_{laser}$) versus RF gun phase scan showing error function dependence on launch phase.  For this scan, setting a fixed $\phi_{gun} = -47\degr$ yields a measurable change in charge for phase fluctuations over a range of a few degrees.}
\end{figure}

We therefore assume an error-function dependence on the charge emitted as a function of gun phase for the purposes of producing a map of charge to phase.  An example of such a phase scan and corresponding fit are shown in Fig.~\ref{fig:phasescan} with the charge emitted normalized to the shot-to-shot laser intensity.

The phase sensitivity will be highest at the center of the rising edge.  For the phase scan shown in Fig.~\ref{fig:phasescan}, the maximum slope of the unnormalized scan is 49.4 pC/degree with a maximum charge of 500 pC.

In our case the charge is monitored by an integrating current transformer (ICT) downstream of the gun.  As noise in the ICT maps to a typical effective time resolution of 1.1 ps RMS ($\sim$0.5$\degr$), this measurement is only sufficient for monitoring drifts of several picoseconds.

\begin{figure}[ht]
\centering
\includegraphics[width=3.05069in]{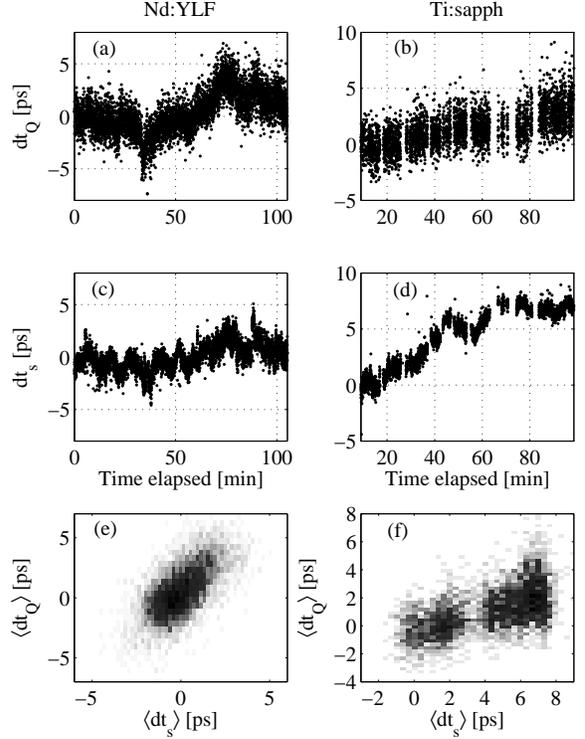}
\caption{\label{fig:jitterplots}Temporal drift data collected with the two techniques for the Nd:YLF (a, c, e) and Ti:sapph (b, d, f) lasers.  The phase difference $dt$ is measured simultaneously using the charge method (a, b) and streak camera (c, d) when the respective laser is used to drive the photoinjector.  For comparison of the drift seen by the two methods, 2D histogram scatter plots of the moving average ${\langle}dt\rangle$ for each pair of data sets are shown below (e, f).}
\end{figure}

Several measurements were performed simultaneously recording data using both methods for each laser with example results shown in Fig.~\ref{fig:jitterplots}.  For both of these sets of plots, the respective laser had a full day of warm up prior to data taking.  Gaps seen in the data for the Ti:sapph system were later found to be caused by improper setup of the 1-kHz \emph{Dazzler} triggering and have since been corrected.

Shot-to-shot fluctuations for all measurements are found to be dominated by the corresponding measurement noise.  For a direct comparison of observed drifts, the moving average of each of the plots is taken to filter out the high frequency jitter.  2D histogram scatter plots of these drifts are shown in Plots (e, f) of Fig.~\ref{fig:jitterplots}.

A weak one-to-one correlation is seen in the data taken for the Nd:YLF system.  For the Ti:sapph, both methods see a similar drift on the scale of a few picoseconds per hour, however this appears exaggerated by the streak camera.  This is likely owing to the streak camera synchroscan unit contributing its own synchronization drift.  In any case, real drifts beyond the stated requirements are observed.

\subsubsection{Simultaneous streaking of two images}
As an alternative to eliminate the ambiguity as regards the streak camera synchronization, simultaneous streaking of two light sources was also performed.  This is particularly useful in comparing the timing of the optical output from the two lasers.  By simultaneously diagnosing the two optical pulses with the same streak camera sweep, the relative phase difference can be measured from shot-to-shot by subtraction thereby eliminating the drift of the images on the screen from the streak camera sweep unit.

\begin{figure}[htb]
\centering
\includegraphics[trim=0cm 1.1cm 0cm 1.2cm, clip=true, width=2.85597in]{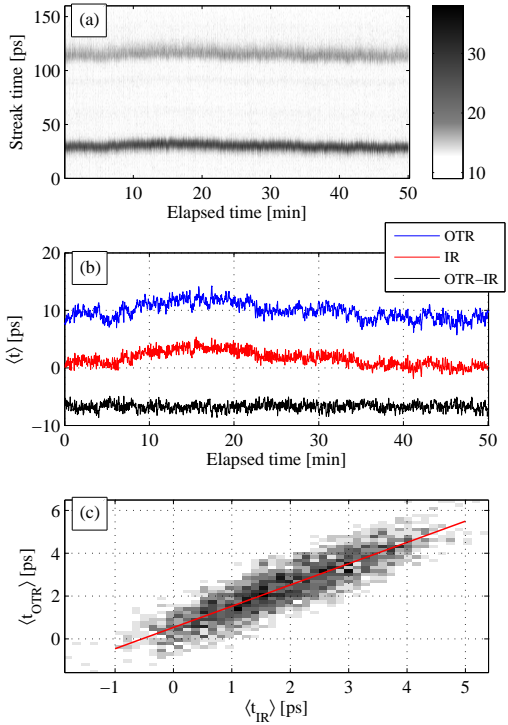}
\caption{\label{fig:irdoublestreak}Streak camera data (sweep range 2) for simultaneous imaging of the Ti:sapph seed laser IR and OTR from the beam using the Ti:sapph system as drive laser.  (a) Raw time-axis projections from the streak camera for the OTR (top) and laser IR (bottom).  (b) Drifts $\langle t \rangle$ deduced from the centroid of double Gaussian fitting after high frequency filtering (arbitrary vertical offsets).  (c) 2D histogram scatter plot of the drifts with inset line of best fit.  Slope of fit line = 0.99.}
\end{figure}

To verify this, Fig.~\ref{fig:irdoublestreak} shows streak data for the Ti:sapph seed laser with OTR from X09 (Fig.~\ref{fig:a0beamline}).  Here the frequency-tripled output of the \emph{Spitfire} is used as drive laser to produce 1.5 nC, 16 MeV bunches.

The synchronization of the 81.25-MHz seed train with the 1-Hz OTR pulse from X09 at the streak camera is achieved using a double-folded retroreflector providing a variable 12 ns optical path delay of the laser.  This is combined with the optical path of the OTR with the delay adjusted to bring both streak images in to view on the same RF sweep.  The time-axis projection of the streak image is recorded shot-to-shot for fitting to determine the temporal centroid of each pulse.

For this measurement, the longer streaks produced in range 1 diluted the photon density of the weak, single-pulse OTR at the streak camera's screen.  To provide a signal measurable over background, sweep range 2 was therefore used instead. The mean streak widths taken from the Gaussian fits to the OTR and IR data in Fig.~\ref{fig:irdoublestreak} were 5.22$\pm$1.63 ps and 2.98$\pm$0.28 ps RMS, respectively.  The deviation in the shot-to-shot phase difference were 2.08 ps and 1.09 ps, respectively.  The larger jitter seen for the OTR track are attributed to larger measurement and fitting noise for the weak OTR signal.

For the nominal operating phase of the RF gun, we expect the gun will not significantly affect the single-particle longitudinal dynamics that would result in drifts from the photocathode laser being mapped onto drifts of the electron bunch time of arrival.  We therefore expect the drifts of the OTR and laser signals to be identical and comprised of the sum of the laser and streak camera synchroscan drifts.

Again applying a high-frequency filter to deduce the drift $\langle t \rangle$, inset (b) of Fig.~\ref{fig:irdoublestreak} includes the difference in drift over time showing effectively zero phase difference within a 0.63 ps RMS spread.   Further, inset (c) shows the scatter plot of the filtered drifts and line of best fit with a slope of 0.99, as expected. Though we cannot decouple the streak camera from laser drift contributions with this information, it reasonably verifies that relative phase measurements can be accomplished with this dual imaging approach with contributions from streak camera PLL drift removed.

\begin{figure}[htb]
\centering
\includegraphics[width=2.92087in]{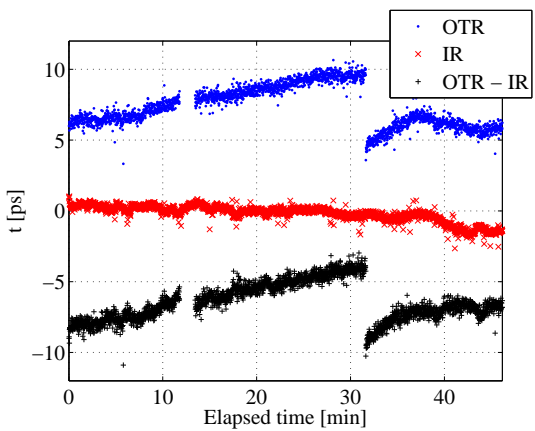}
\caption{\label{fig:irandylfotr}Relative time of arrival data from simultaneous (sweep range 1) streaking of the Ti:sapph IR and OTR generated by beam produced using the Nd:YLF system as drive laser (arbitrary vertical offsets).}
\end{figure}

As relates to phase-locking between lasers for EOSD, we repeat this experiment using the Ti:sapph IR and OTR from the Nd:YLF-driven electron beam.  Several sweeps of the streak camera can then be stacked to produce a stronger signal allowing for fast, range 1 measurements assuming negligible phase drift between the 1 MHz-spaced bunches.  Results are shown in Fig.~\ref{fig:irandylfotr} using 80, 500-pC bunches with the streak camera operating in its fastest, range 1 sweep mode.  The 2-min gap in the data for the OTR at 12 min was due to temporary loss of gun RF.

For this set, the streak widths for the IR and OTR were 1.11$\pm$0.05 ps and 4.12$\pm$.16 ps RMS, respectively.  Shot-to-shot jitter was 160 fs and 434 fs, respectively, with the larger value for the OTR signal attributed to the longer streak length leading to a greater inherent measurement uncertainty.

We note in Fig.~\ref{fig:irandylfotr} an 8.5 ps/hour relative drift between the two signals manifesting primarily in the OTR signal as well as a 5.5 ps phase jump at 32 min.  This has been identified as a signature of the Nd:YLF seed laser's phase lock loop.  A frequency difference between the seed cavity's 81.25 MHz and reference to the master oscillator is erroneously fed back into the cavity with the piezo experiencing a relatively constant displacement drift until reaching its limit of travel.  At this point the electronics adjust the picomotor stage upon which the piezo mirror is mounted to reset its position to the center of the piezo travel.  This movement of the stage causes the phase jump in the seed output.

The periodicity of these resets and the slope of the drift between them have been found to be highly sensitive to the alignment, optical power and warm up of the Nd:YLF oscillator making consistent correction by laser tune up alone difficult.

We observe that the spread from drift contributions in these data sets account for the significant, ps-scale portion of timing instability with drifts exceeding 5 ps/hour observed in the Nd:YLF system.

\subsubsection{Secondary laser phase monitor and feedback}\label{sec:secondarypll}
\begin{figure}[htb]
\centering
\includegraphics[width=3.05069in]{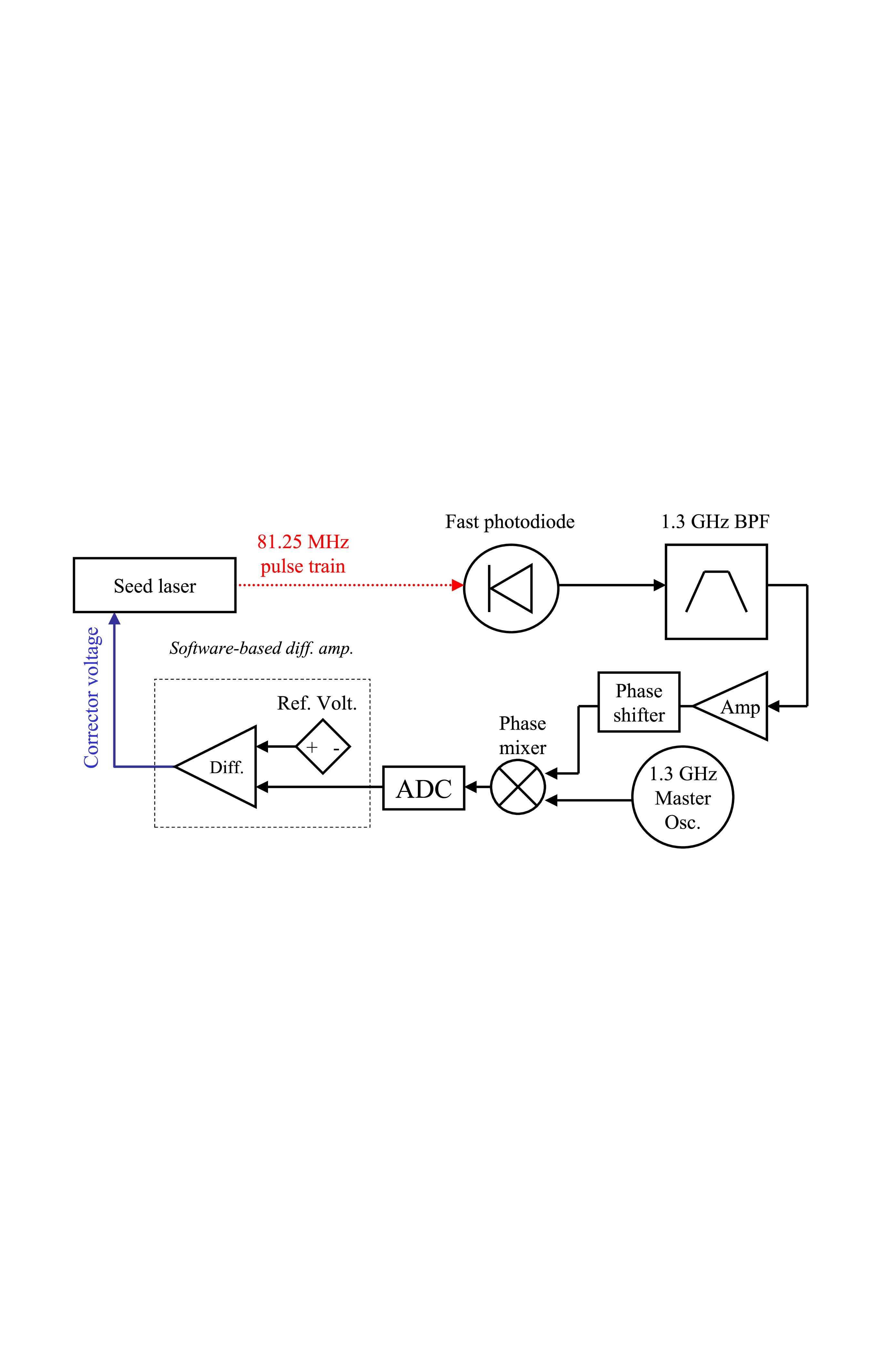}
\caption{\label{fig:feedbackloop}Block diagram of the laser phase monitor and software-based feedback used to compensate drift in both seed lasers.}
\end{figure}

To correct the ps-scale drift of the seed lasers while also providing an additional measure of laser phase stability with respect to the RF, we use a setup similar to that demonstrated in~\cite{fssync1, fssync2}, shown in Fig.~\ref{fig:feedbackloop}.  For phase detection, optical leakage is sent to a fast photodiode.  The resulting signal then passes a 1.3 GHz band-pass cavity filter and low-noise amplifier.  The phase is compared to that of the 1.3-GHz master oscillator in a phase mixer with input RF levels attenuated to the operational range of the mixer and a variable phase delay on the diode signal so the phase difference can be set to operate in the mixer's linear response region.

Mixer output can be used to monitor the phase difference between the laser and master with a conversion factor of 100 mV/deg.  A 4-ms sample is converted by an analog-to-digital converter (ADC) once per second with a sampling rate of 1 MHz.  The spread of the sample is recorded to monitor jitter and has resolved noise as low as 200 fs RMS while the mean voltage of each sample is tracked for phase drift.

The synchronization electronics of both systems include input ports to which an external bias can be applied that's added into the signal driving the piezo-mounted mirrors.  This is employed in feedback mode using software-based differential amplification of the ADC output.  The mean of the sample is compared to a set reference voltage and, with the appropriate programmable gain, generates a corrector voltage sent back to the associated seed laser via an internet rack monitor (IRM)~\cite{irm} signal.

The digital IRM channels used to provide the programmable corrector voltage have a minimum step size of 5 mV.  For the Ti:sapph system, this was found to be too large to provide adequate time resolution given the feedback sensitivity of the electronics.  To correct this, a 1/9 analog voltage divider is added to the phase input port of the \emph{Tsunami} to reduce the effective minimum step size of the IRM output to acceptable levels with changes to the effective differential amplifier gain accounted for.

\subsubsection{Measurement of corrected laser-to-laser drift}
\begin{figure}[htb]
\centering
\includegraphics[width=3.05069in]{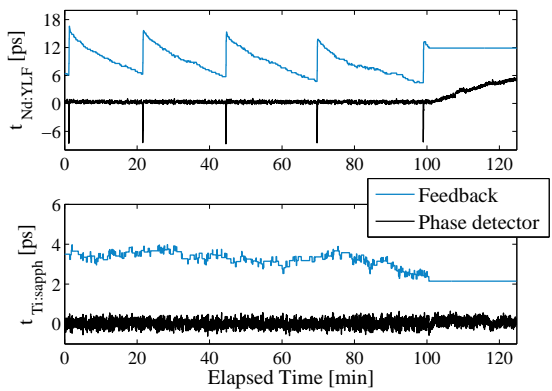}
\caption{\label{fig:feedbackplot}Time of arrival data taken simultaneously for the Nd:YLF (top) and Ti:sapph (bottom) laser systems.  The feedback corrector signals (blue) are generated in response to the RF phase detectors (black) with feedback disabled at 100 min.}
\end{figure}

Feedback loop performance for phase stability between the two lasers was verified by again analyzing simultaneous streak imaging of the Nd:YLF UV and amplified Ti:sapph IR outputs.  Results for phases tracked by the phase detector in the feedback loop versus that on the streak camera are compared in Fig.~\ref{fig:feedbackplot}.  The phase as recorded by the streak camera and loops are plotted with the concurrent secondary feedback being applied with feedback disabled at 100 min.

Prior to this measurement, the Ti:sapph system was warmed up over a full day while in use for other experiments.  As such, the corrector phase being generated for the Ti:sapph laser is relatively stable to the order of the laser jitter.  In fact, the spread of the long-term projection seen on the phase detector is the same ($<$ 200 fs, RMS) whether or not feedback is enabled in this set.

\begin{figure}[htb]
\centering
\includegraphics[width=3.24542in]{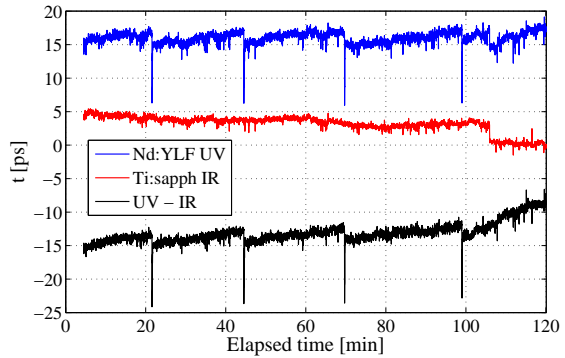}
\caption{\label{fig:iranduvfixed}Time of arrival $t$ determined from simultaneous streak camera imaging for Ti:sapph (red) and Nd:YLF (blue) lasers and the deduced time difference (black).  Both lasers are imaged to the slit of the synchronized streak camera in the accelerator tunnel.  Data taken concurrently with that shown in Fig.~\ref{fig:feedbackplot} with feedback enabled from 0--100 min.  }
\end{figure}

The relatively stable output is nonetheless shown here for comparison to the Ti:sapph output as measured on the streak camera, shown in Fig.~\ref{fig:iranduvfixed}.  We observe that the streak camera data shows an abrupt change in the Ti:sapph streak image at 106 minutes that does not appear on the phase monitor which is also the case for the Nd:YLF measurements.  This indicates a disruption in the PLL for the streak camera which is indeed removed after taking the difference in the output phases from the streak camera measurement (Fig.~\ref{fig:iranduvfixed}).

In the data for the Nd:YLF system (Fig.~\ref{fig:feedbackplot}), the clear sawtooth pattern as described earlier is observed in the feedback signal.  The system tracks several steep drifts with phase jumps of 9.4 ps spaced roughly every 30 min with the periodicity growing a few minutes per cycle.  The corresponding output phase as measured by the phase detector appears flat with the spread again on the order of the laser jitter, excepting the few cycles it takes to recover the phase lock after a jump.

The magnitude of these changes are also observable on the streak camera (Fig.~\ref{fig:iranduvfixed}) over the few cycles it takes the feedback to recover the phase.  However, the streak camera reveals that the secondary phase lock loop is not fully correcting the drift with the discrete phase jumps instead appearing with a magnitude of 11.2 ps.  As a result, a 1.8 ps-amplitude sawtooth drift survives in the difference signal where the few-cycle disruption from the discrete changes are neglected.

With secondary feedback disabled at 100 min, the relative phase immediately begins to run beyond this 1.8 ps drift oscillation.  With feedback, however, and excluding extreme outliers from the Nd:YLF laser phase jumps, the spread of all points taken with loops enabled is 0.81 ps RMS.  This brings us within the desired sub-ps specification for long-term timing stability.
\section{Temporal pulse shape measurement}
To verify the temporal pulse shape of the output laser pulse for the associated experiment, FROG using second harmonic generation (SHG FROG) was used~\cite{frog1, frog2} for complete laser phase reconstruction.

For EOSD this gives the information required to map wavelength to time for decoding the signal sampled from the electron beam.  For the blow-out experiment, it's to verify the laser pulse is as near bandwidth-limited as possible to generate the shortest UV pulse after tripling.  Information about the spectral phase can be fed back into the \emph{Dazzler} to correct third order dispersion (TOD), the third coefficient in the Taylor expansion of the spectral phase, from the cavity for more efficient tripling and a more uniform UV pulse.

The SHG FROG optical hardware is built on a kit from Newport Corp. (part no. FRG-KT)~\cite{newport} using a 200~$\muup$m BBO crystal cut for 800-nm frequency doubling.  The supplied spectrometer was replaced with an Ocean Optics \emph{Jaz} spectrometer with an 1800 lines/mm grating.  FROG traces are analyzed using the \emph{FROG} software package by Femtosoft Technologies~\cite{frog3}.  A MATLAB interface manages data acquisition and formatting for input and output with the reconstruction software.

For EOSD, a pulse length of as much as 5 ps is desired to allow a sufficiently long sampling window.  As the \emph{Dazzler} can only produce a maximum chirped length of approximately 2 ps for the given laser bandwidth, the grating compressor of the \emph{Spitfire} was instead adjusted to provide the longer pulse.  Moderate variations to the nominal pulse length can then be made on the fly using the programmable setting of the \emph{Dazzler}.  As the compressor stage of the \emph{Spitfire} provides no readback for pulse length or stage position settings, independent measurement of a chosen setting is also done by SHG FROG.

\begin{figure}[htbp]
\centering
\includegraphics[width=3.1156in]{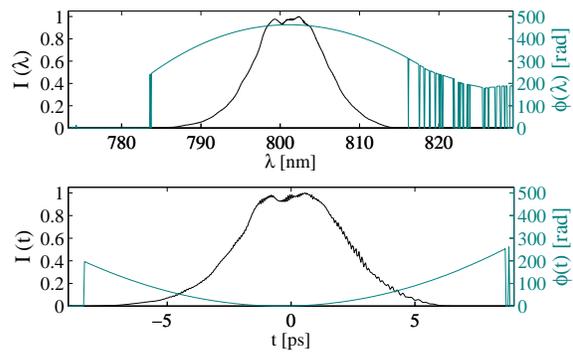}
\caption{\label{fig:chirped}Reconstructed spectral (top) and temporal (bottom) profiles and phases for a strongly chirped laser pulse with a spectral bandwidth of 10.5 nm and pulse duration of 4.9 ps, FWHM.}
\end{figure}

An example chirped pulse measurement is shown in Fig.~\ref{fig:chirped} showing the reconstructed amplitude and phase information for the pulse.  For diagnosing a long pulse such as this, a large $1024\times1024$ pixel FROG grid must be used to satisfy the Nyquist conditions as the bandwidth demands a time step much smaller than the large relevant time scale.  The associated FROG error was 0.798\% yielding a bandwidth of 10.5 nm FWHM with group delay dispersion of $1.61\times10^5$ fs$^2$ and corresponding pulse width of 4.9 ps FWHM.  The detailed phase information provided by this measurement alleviates errors in the decoding of EOSD signals as the laser pulse acts as the carrier upon which the diagnostic signal is encoded.

For ultrashort pulse generation we start by measuring the spectral phase of the maximally compressed \emph{Spitfire} output.  To reduce the effect of statistical error, the FROG measurement and reconstruction are repeated ten times with the \emph{Dazzler} operating in self-compensating mode.  The average of the recovered spectral phase curves is taken as their mean after setting a constant phase and slope at $\lambda$ = 800 nm for all traces to neglect the known ambiguity from SHG FROG reconstruction.  This phase curve is then programmed for subtraction by the \emph{Dazzler} and the measurement is repeated.

\begin{figure}[htbp]
\centering
\includegraphics[width=3.1156in]{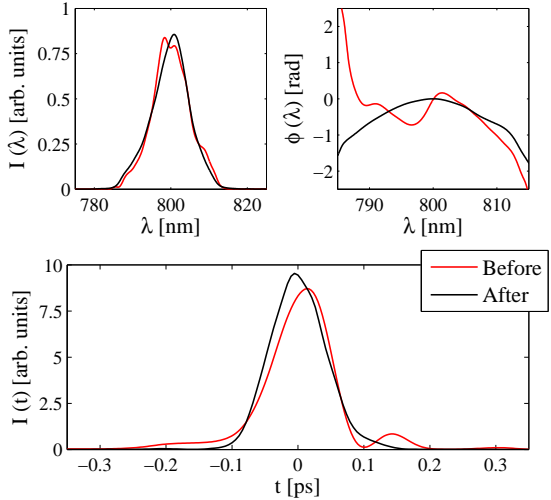}
\caption{\label{fig:compensate}Reconstructed temporal and spectral profiles of a maximally compressed pulse before and after compensating for the measured third order dispersion.}
\end{figure}

The result is shown in Fig~\ref{fig:compensate} for the pulse pre- and post-compensation.  The FROG traces shown here all fit with an error of $<$ 0.35\%.  Before adding TOD compensation a cubic spectral phase with TOD of greater than $6\times10^5$ fs$^3$ is observed producing a long tail in the temporal profile.  With the measured spectral phase subtracted by the \emph{Dazzler} pulse shaper, the shoulders in the field are suppressed though a small, 827 fs$^2$ of second order dispersion remains.  The initial profile standard deviation of 78.7 fs is reduced to 49.2 fs resulting in a uniform IR pulse with a 98.4 fs intensity FWHM.

\section{Future Improvements}
Detailed investigation of the remaining phase drift after secondary feedback (Fig.~\ref{fig:iranduvfixed}) has not yet been explored.  Phase/frequency modulation imparted by the photodiodes and digitization error introduced by the software-based loops are suspected.

A higher bandwidth (1 kHz), fully analog feedback system is also available that may mitigate problems arising in the ADC and IRM.  In this case, the output of the phase mixer (Fig.~\ref{fig:feedbackloop}) is passed to an analog, fixed-gain differential amplifier to produce the corrector voltage.  The software solution is currently still used as it offers programmable parameters and machine protection features not present in the analog design.  The aggressive feedback provided by the analog amplifier in its present state was found to be intolerant of incidental perturbations in the mixer signal with the potential of over-driving the corrector voltage and, therefore, the seed laser piezo.  As a result the phase lock would on occasion become unstable after less than an hour of operation.


\section{Summary}
A commercial Ti:sapph laser and transport optics have been installed and successfully commissioned at the A0 photoinjector laboratory at Fermilab.  The system is reliably synchronized to the 1-Hz RF pulse generation and instrumentation trigger with the seed laser exhibiting temporal jitter of less than 300 fs RMS.

The long-term phase stability of both the existing Nd:YLF drive laser and new Ti:sapph system has been diagnosed by a number of independent experiments to allow diagnosis of the Nd:YLF-driven beam by EOSD using the Ti:sapph as probe laser.  Simultaneous long-term synchronization of the seeds to within 1 ps of the 1.3-GHz master clock is accomplished using independent secondary feedback loops available for both systems.

Intensity stability of amplified IR output is found to be 1.5\% RMS with a corresponding UV pulse stability of 4\%.  An SHG FROG has been assembled for detailed temporal laser phase space reconstruction allowing ultrashort pulse optimization and laser chirp characterization.  At maximum compression pulse lengths of 100 fs FWHM are produced, sufficient to drive ellipsoidal bunch generation studies.  Chirped pulse lengths of 5 ps can be easily set and measured for applications in EOSD experiments.   The laser system has subsequently been used to successfully produce ellipsoidal bunches~\cite{a0ellipse} and support single-shot measurements of the electron bunch duration downstream of the emittance-exchange beamline~\cite{myeos}.

\section*{Acknowledgements}
We would like to thank Michael Kucera and James Santucci of Fermi National Accelerator laboratory, all of our colleagues with the A0 photoinjector group, as well as Michael Maikowski and Art Camire at Newport Corporation's Spectra Physics division for helpful discussions and technical support.  This work was supported by the Fermi Research Alliance, LLC under U.S. Department of Energy Contract Number DE-AC02-07CH11359, and Northern Illinois University under U.S. Department of Defense DURIP program Contract N00014-08-1-1064.

\bibliographystyle{model1-num-names}
\bibliography{Maxwell_NIMA_manuscript}{}

\end{document}